\documentclass[conference,10pt]{IEEEtran}

\usepackage{cite}

\ifCLASSINFOpdf
\else
   \usepackage[dvipdfmx]{graphicx}
\fi
\usepackage[cmex10]{amsmath}
\usepackage{amssymb}
\usepackage{balance}

\newtheorem{theorem}{Theorem}
\newtheorem{lemma}{Lemma}

\newtheorem{assumption}{Assumption}
\newtheorem{remark}{Remark}

\begin{document}
%
\title{Asymptotic Optimality of Massive MIMO Systems Using Densely Spaced Transmit Antennas}
%
%
%

\author{\IEEEauthorblockN{Keigo Takeuchi}
\IEEEauthorblockA{Dept.\ Commun.\ Eng.\ \& 
Informatics, University of Electro-Communications, 
Tokyo 182-8585, Japan}
\IEEEauthorblockA{Email: ktakeuchi@uec.ac.jp} 
}

\maketitle

\begin{abstract}
This paper investigates the performance of a massive multiple-input 
multiple-output (MIMO) system that uses a large transmit antenna array with 
antenna elements spaced densely. Under the assumption of idealized 
uniform linear antenna arrays without mutual coupling, precoded 
quadrature phase-shift keying (QPSK) transmission is proved to achieve 
the channel capacity of the massive MIMO system when the transmit antenna 
separation tends to zero. This asymptotic optimality is analogous to that 
of QPSK faster-than-Nyquist signaling.  
\end{abstract}


%

\section{Introduction}
Half wavelength was proved to be the critical antenna separation to attain 
the maximum spatial degrees of freedom in multiple-input multiple-output 
(MIMO) systems~\cite{Poon05}. Thus, there are no points in considering 
more densely spaced antennas when optimal Gaussian signaling is used. 
The purpose of this paper is to investigate the impact of transmit antenna 
separation on achievable rate when suboptimal quadrature phase-shift keying 
(QPSK) is used. 

We exploit an analogy between a transmit antenna array and a band-limited 
system, pointed out in \cite{Poon05}. Let $L_{\mathrm{t}}$ denote the array 
length normalized by the carrier wavelength. The spatial domain 
$[-L_{\mathrm{t}}/2, L_{\mathrm{t}}/2]$ in the array corresponds to the time domain 
in the band-limited system with bandwidth $W=2$, while the angular domain 
$[-1 ,1]$ does to the frequency domain $[-W/2, W/2]$. The critical 
normalized antenna separation $\Delta_{\mathrm{t}}=1/2$ is due to Nyquist's 
sampling theorem, claiming that sampling at the Nyquist period $1/W$ 
reconstructs any band-limited signal, when the array length $L_{\mathrm{t}}$ 
tends to infinity. 

Faster-than-Nyquist (FTN) signaling~\cite{Mazo75,Rusek09} increases the 
transmission rate of band-limited systems by sending pulses at a symbol 
period shorter than the Nyquist period. Yoo and Cho~\cite{Yoo10} proved 
that QPSK FTN signaling achieves the channel capacity of the additive white 
Gaussian noise (AWGN) channel when the symbol period tends to zero. 
The main contribution of this paper is to prove an analogous result: 
Precoded QPSK transmission can achieve the channel capacity of massive MIMO 
systems when the transmit antenna separation tends to zero. 

\section{System Model}
\subsection{MIMO Channel}
Consider a MIMO channel with $M$ transmit antennas and $N$ receive antennas. 
The received vector $\boldsymbol{y}\in\mathbb{C}^{N}$ is given by 
\begin{equation} \label{MIMO}
\boldsymbol{y} = \sqrt{\gamma}\boldsymbol{H}\boldsymbol{x} 
+ \boldsymbol{w},
\end{equation}
with $\boldsymbol{w}\sim\mathcal{CN}(\boldsymbol{0},\boldsymbol{I}_{N})$. 
In (\ref{MIMO}), $\boldsymbol{H}\in\mathbb{C}^{N\times M}$ denotes the channel 
matrix between the transmitter and the receiver. The vector 
$\boldsymbol{x}\in\mathbb{C}^{M}$ is the transmitted vector with power 
constraint $\mathbb{E}[\|\boldsymbol{x}\|^{2}]\leq 1$, so that the parameter 
$\gamma>0$ corresponds to the signal-to-noise ratio (SNR). 

\subsection{Uniform Linear Antenna Array}
In order to introduce a deterministic physical model of the channel matrix, 
we consider idealized uniform linear antenna arrays without mutual coupling. 
In realistic antenna arrays, coupling between adjacent antenna elements occurs 
when the antenna elements are spaced closely~\cite{Janaswamy02}. 
However, the influence of mutual coupling may be mitigated by constructing a 
matching network at each antenna array~\cite{Wallace04}. For simplicity, 
the influence of mutual coupling is ignored in this paper. 

Let $\Delta_{\mathrm{t}}\in(0, 1/2]$ and $\Delta_{\mathrm{r}}\in(0, 1/2]$ denote 
the transmit and receive antenna separations normalized by the carrier 
wavelength, respectively, while the normalized lengths of the transmit and 
receive antenna arrays are written as $L_{\mathrm{t}}$ and $L_{\mathrm{r}}$. 
We assume that each array length is a multiple of the corresponding antenna 
separation, so that $N=L_{\mathrm{r}}/\Delta_{\mathrm{r}}$ and 
$M=L_{\mathrm{t}}/\Delta_{\mathrm{t}}$ hold. Then, we follow 
\cite[Eq.~(7.56)]{Tse05} to model 
the channel matrix $\boldsymbol{H}$ as follows:
\begin{equation} \label{modeling}
\boldsymbol{H} 
= \sqrt{NM}\int a(p)\boldsymbol{s}_{L_{\mathrm{r}}, \Delta_{\mathrm{r}}}
(\Omega_{\mathrm{r}}(p))\boldsymbol{s}_{L_{\mathrm{t}}, \Delta_{\mathrm{t}}}^{\mathrm{H}}
(\Omega_{\mathrm{t}}(p))dp. 
\end{equation}

In (\ref{modeling}), $a(p)$ denotes the complex attenuation of path~$p$. 
The directional cosines $\Omega_{\mathrm{t}}(p)=\cos\phi_{\mathrm{t}}(p)$ and 
$\Omega_{\mathrm{r}}(p)=\cos\phi_{\mathrm{r}}(p)$ for path $p$ are defined via the 
departure angle $\phi_{\mathrm{t}}(p)$ from the transmit antenna array and via 
the incidence angle $\phi_{\mathrm{r}}(p)$ to the receive antenna array, 
respectively. For a uniform linear antenna array with normalized array length 
$L$ and antenna separation $\Delta$, the normalized spatial signature vector 
$\boldsymbol{s}_{L,\Delta}(\Omega)\in\mathbb{C}^{L/\Delta}$ with respect to 
directional cosine $\Omega$ is given by 
\begin{equation} \label{signature}
\boldsymbol{s}_{L,\Delta}(\Omega)
= \frac{1}{\sqrt{L/\Delta}}
\left(
 1, e^{-2\pi j\Delta\Omega}, \ldots, e^{-2\pi j(L/\Delta -1)\Delta\Omega}
\right)^{\mathrm{T}}, 
\end{equation}
where $j$ denotes the imaginary unit. 
 
\subsection{Angular Domain Representation}
We next introduce the angular domain representation of the channel 
matrix~(\ref{modeling}). Since 
$\{\boldsymbol{s}_{L,\Delta}(k/L) | k\in[0 : L/\Delta)\}$\footnote{
For integers $a$ and $b$, $[a: b)$ represents the 
set $\{a, a+1,\ldots, b-1\}$. 
} form an orthonormal basis of $\mathbb{C}^{L/\Delta}$~\cite{Tse05}, 
we can represent any signature vector $\boldsymbol{s}_{L,\Delta}(\Omega)$ as 
\begin{equation} \label{expansion}
\boldsymbol{s}_{L,\Delta}(\Omega) 
= \sum_{k=0}^{L/\Delta-1}f_{L,\Delta}\left(
 \frac{k}{L} - \Omega
\right)\boldsymbol{s}_{L,\Delta}\left(
 \frac{k}{L}
\right),
\end{equation} 
with 
\begin{equation}
f_{L,\Delta}(\Omega) 
= \boldsymbol{s}_{L,\Delta}^{\mathrm{H}}(0)\boldsymbol{s}_{L,\Delta}(\Omega). 
\end{equation}
In the derivation of (\ref{expansion}), we have used the fact that the 
inner product 
$\boldsymbol{s}_{L,\Delta}^{\mathrm{H}}(\Omega')\boldsymbol{s}_{L,\Delta}(\Omega)$ 
depends only on the difference $\Omega-\Omega'$. Applying (\ref{expansion}) 
to the channel matrix~(\ref{modeling}) yields 
\begin{equation} \label{angular_domain}
\sqrt{\gamma}\boldsymbol{H} 
= \sqrt{\tilde{\gamma}}\boldsymbol{U}_{L_{\mathrm{r}},\Delta_{\mathrm{r}}}
\boldsymbol{G}_{\Delta_{\mathrm{t}},\Delta_{\mathrm{r}}}
\boldsymbol{U}_{L_{\mathrm{t}},\Delta_{\mathrm{t}}}^{\mathrm{H}},
\end{equation} 
where we have defined the normalized SNR 
$\tilde{\gamma}=\gamma/(4\Delta_{\mathrm{t}}\Delta_{\mathrm{r}})$. 

In (\ref{angular_domain}), the $k$th column of the $L/\Delta\times L/\Delta$ 
unitary matrix $\boldsymbol{U}_{L,\Delta}$ is given by 
$\boldsymbol{s}_{L,\Delta}(k/L)$ for all $k\in[0: L/\Delta)$, so that 
$\boldsymbol{U}_{L,\Delta}$ is the $L/\Delta$-point discrete Fourier 
transform (DFT) matrix. Furthermore, the $(n, m)$ entry $g_{n,m}$ of the 
channel matrix 
$\boldsymbol{G}_{\Delta_{\mathrm{t}},\Delta_{\mathrm{r}}}\in\mathbb{C}^{N\times M}$ 
in the angular domain is given by 
\begin{IEEEeqnarray}{r}
g_{n,m}
= \sqrt{4L_{\mathrm{t}}L_{\mathrm{r}}}\int a(p)
f_{L_{\mathrm{r}},\Delta_{\mathrm{r}}}\left(
 \frac{n}{L_{\mathrm{r}}} - \Omega_{\mathrm{r}}(p)
\right) 
\nonumber \\ 
\cdot f_{L_{\mathrm{t}},\Delta_{\mathrm{t}}}^{*}\left(
 \frac{m}{L_{\mathrm{t}}} - \Omega_{\mathrm{t}}(p)
\right)dp. 
\end{IEEEeqnarray}

\begin{remark}
The normalized SNR $\tilde{\gamma}$ in (\ref{angular_domain}) increases 
as the antenna separation $\Delta_{\mathrm{t}}$ decreases. This improvement 
in SNR results from the power gain obtained by using multiple 
antennas~\cite{Tse05}. However, we have ignored an effective power loss 
due to mutual coupling, so that it may be fair to compare different systems 
with identical normalized SNR $\tilde{\gamma}$. 
\end{remark}

\subsection{Constrained Capacity} 
Let $\boldsymbol{Q}_{M}$ denote an $M\times M$ input covariance matrix with 
power constraint $\mathrm{Tr}(\boldsymbol{Q}_{M})\leq1$. It is well known 
that the constrained capacity $C_{\mathrm{opt}}(\boldsymbol{Q}_{M};
\tilde{\gamma}, \boldsymbol{G}_{\Delta_{\mathrm{t}},\Delta_{\mathrm{r}}})$ of the MIMO 
channel~(\ref{MIMO}) with the channel matrix~(\ref{angular_domain}) is given by 
\begin{IEEEeqnarray}{rl} 
&C_{\mathrm{opt}}(\boldsymbol{Q}_{M};
\tilde{\gamma}, \boldsymbol{G}_{\Delta_{\mathrm{t}},\Delta_{\mathrm{r}}})  
\nonumber \\ 
=& \log\det\left(
 \boldsymbol{I}  
 + \tilde{\gamma}\boldsymbol{G}_{\Delta_{\mathrm{t}},\Delta_{\mathrm{r}}}
 \boldsymbol{U}_{L_{\mathrm{t}},\Delta_{\mathrm{t}}}^{\mathrm{H}}
 \boldsymbol{Q}_{M}
 \boldsymbol{U}_{L_{\mathrm{t}},\Delta_{\mathrm{t}}}
 \boldsymbol{G}_{\Delta_{\mathrm{t}},\Delta_{\mathrm{r}}}^{\mathrm{H}}
\right), \label{capacity}
\end{IEEEeqnarray}
which is achieved by Gaussian signaling 
$\boldsymbol{x}\sim\mathcal{CN}(\boldsymbol{0},\boldsymbol{Q}_{M})$. 

The channel capacity is the maximum of the constrained 
capacity~(\ref{capacity}) over all possible input covariance matrices 
$\boldsymbol{Q}_{M}$. Poon {\em et al.}~\cite{Poon05,Poon06} proved that 
the maximum spatial degrees of freedom in the MIMO channel are equal to 
$2\mathrm{min}\{L_{\mathrm{t}},L_{\mathrm{r}}\}$: The capacity is proportional 
to $2\mathrm{min}\{L_{\mathrm{t}},L_{\mathrm{r}}\}\log\tilde{\gamma}$ in the 
high SNR regime under richly scattering environments. 

\section{Main Results}
\subsection{Large-System Analysis}
Large-system analysis has been considered to evaluate the performance of 
massive MIMO systems, on the basis of random matrix 
theory~\cite{Tse99,Verdu99,Tulino05} or of the replica method~\cite{Tanaka02,Moustakas03,Takeda06,Wen07,Takeuchi08,Mueller08,Girnyk14}. Since statistical 
channel models were used in previous works, the system parameters were 
the numbers of transmit and receive antennas. In our deterministic model, 
however, the array length and the antenna separation are the fundamental 
parameters to describe the channel model~\cite{Poon05}. Thus, we consider the 
large-system limit in which $M$, $N$, $L_{\mathrm{t}}$, and $L_{\mathrm{r}}$ tend 
to infinity while the ratios $\Delta_{\mathrm{t}}=L_{\mathrm{t}}/M$, 
$\Delta_{\mathrm{r}}=L_{\mathrm{r}}/N$, and $\alpha=L_{\mathrm{t}}/L_{\mathrm{r}}$ are 
kept constant. This is the precise meaning of massive MIMO in this paper. 

In this paper, we discuss uniform convergence over a class $\mathfrak{C}$ 
of channel instances $\mathcal{C}=\{a(\cdot), \Omega_{\mathrm{t}}(\cdot), 
\Omega_{\mathrm{r}}(\cdot)\}$. 
\begin{assumption} \label{assumption1}
Let $\{s_{l}\in\mathbb{N}\}_{l=1}^{\infty}$ denote a slowly diverging sequence of 
natural numbers, which means the limits $s_{l}\to\infty$ and $s_{l}/l\to0$ 
as $l\to\infty$. For $\mathcal{D}_{l}=[-(1-s_{l}/l), 1-s_{l}/l]$, postulate 
a class $\mathfrak{C}$ of channel instances (a set of $\mathcal{C}$) such 
that i) $\Omega_{\mathrm{t}}(\cdot)\in\mathcal{D}_{L_{\mathrm{t}}}$ and 
$\Omega_{\mathrm{r}}(\cdot)\in\mathcal{D}_{L_{\mathrm{r}}}$ hold, and that ii) 
the total power $\int|a(p)|^{2}dp$ of attenuation and the maximum singular 
value of $\mathrm{min}\{2L_{\mathrm{t}},2L_{\mathrm{r}}\}^{-1/2}
\boldsymbol{G}_{\Delta_{\mathrm{t}},\Delta_{\mathrm{r}}}$ are uniformly bounded in the 
large-system limit for all channel instances $\mathcal{C}\in\mathfrak{C}$. 
\end{assumption}

The latter assumption forbids one to enjoy noise-less eigen channels 
in the large-system limit. The former assumption is a sufficient condition 
for proving the main theorems with respect to Gaussian signaling. Note that 
this assumption should almost surely hold in the large-system limit, if channel 
instances are sampled from proper statistical models. 

\subsection{Gaussian Signaling}
For any input covariance matrix $\boldsymbol{Q}_{M}$, we define the covariance 
matrix $\boldsymbol{\Sigma}_{M}
=\boldsymbol{U}_{L_{\mathrm{t}},\Delta_{\mathrm{t}}}^{\mathrm{H}}
\boldsymbol{Q}_{M}\boldsymbol{U}_{L_{\mathrm{t}},\Delta_{\mathrm{t}}}$
to specify the structure of the optimal covariance matrix in the large-system 
limit. 

\begin{figure}[t]
\begin{center}
\includegraphics[width=\hsize]{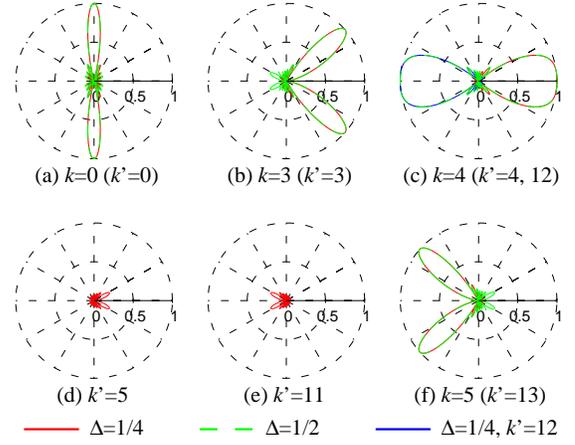}
\caption{
Beamforming patterns $(\phi, |f_{L,\Delta}(k/L-\cos\phi)|)$ for the critically 
spaced case $\Delta=1/2$, and $(\phi, |f_{L,\Delta}(k'/L-\cos\phi)|)$ 
for the densely spaced case $\Delta=1/4$. The normalized array length is set 
to $L=4$.   
}
\label{fig1}
\end{center}
\end{figure}

When the densely spaced case $\Delta_{\mathrm{t}}, \Delta_{\mathrm{r}}<1/2$ is 
considered, as indicated from Fig.~\ref{fig1}, most of the power of the 
channel gains $g_{n,m}$ is concentrated on indices 
$n\in\mathcal{N}=[0:L_{\mathrm{r}}]\cup
[N-L_{\mathrm{r}}: N)$ and $m\in\mathcal{M}=[0:L_{\mathrm{t}}]\cup
[M-L_{\mathrm{t}}: M)$ in the large-system limit. In order to present a 
precise statement on this observation, we shall define the effective channel 
matrix $\tilde{\boldsymbol{G}}_{\Delta_{\mathrm{t}},\Delta_{\mathrm{r}}}$ as 
\begin{equation} \label{tilde_G}
(\tilde{\boldsymbol{G}}_{\Delta_{\mathrm{t}}, \Delta_{\mathrm{r}}})_{n,m} 
= \left\{
\begin{array}{cl}
g_{n,m} & \hbox{for $(n, m)\in\mathcal{N}\times\mathcal{M}$,} \\ 
0 & \mathrm{otherwise.}
\end{array}
\right.
\end{equation}

\begin{assumption} \label{assumption2}
Postulate the set $\mathfrak{S}_{M}$ of $M\times M$ covariance matrices 
such that, for all $\boldsymbol{\Sigma}_{M}\in\mathfrak{S}_{M}$, the power 
constraint $\mathrm{Tr}(\boldsymbol{\Sigma}_{M})\leq1$ is satisfied, and that
the maximum eigenvalue of $2L_{\mathrm{t}}\boldsymbol{\Sigma}_{M}$ is 
uniformly bounded in the large-system limit. 
\end{assumption}

\begin{theorem} \label{theorem1} 
Let $\boldsymbol{Q}_{M}=\boldsymbol{U}_{L_{\mathrm{t}},\Delta_{\mathrm{t}}}
\boldsymbol{\Sigma}_{M}\boldsymbol{U}_{L_{\mathrm{t}},\Delta_{\mathrm{t}}}^{\mathrm{H}}$, and 
fix SNR $\tilde{\gamma}>0$, receive antenna separation 
$\Delta_{\mathrm{r}}\in(0,1/2]$, and load $\alpha=L_{\mathrm{t}}/L_{\mathrm{r}}>0$. 
Under Assumptions~\ref{assumption1} and \ref{assumption2}, 
\begin{equation} \label{asym_capacity}
\frac{|C_{\mathrm{opt}}(\boldsymbol{Q}_{M};\tilde{\gamma}, 
 \boldsymbol{G}_{\Delta_{\mathrm{t}}, \Delta_{\mathrm{r}}})
 - C_{\mathrm{opt}}(\boldsymbol{Q}_{M};\tilde{\gamma}, 
 \tilde{\boldsymbol{G}}_{\Delta_{\mathrm{t}}, \Delta_{\mathrm{r}}})|
}{2\min\{L_{\mathrm{t}},L_{\mathrm{r}}\}} \to0
\end{equation}
holds uniformly for all transmit antenna separations 
$\Delta_{\mathrm{t}}\in(0,1/2]$,  
covariance matrices $\boldsymbol{\Sigma}_{M}\in\mathfrak{S}_{M}$, 
and all channel instances $\mathcal{C}\in\mathfrak{C}$ in the large-system 
limit. 
\end{theorem}
\begin{IEEEproof}
See a long version of this paper~\cite{Takeuchi16}. 
\end{IEEEproof}

Theorem~\ref{theorem1} implies that, as long as the large-system limit 
is taken, it is sufficient to consider covariance matrices 
$\boldsymbol{\Sigma}_{M}$ with all-zero vectors in columns 
$(L_{\mathrm{t}}:M-L_{\mathrm{t}})$ and in rows $(L_{\mathrm{t}}:M-L_{\mathrm{t}})$. 
Under this restriction, we next 
make a comparison between the critically spaced case and the densely spaced 
case. For notational convenience, we define 
$\boldsymbol{E}_{n,k}(\boldsymbol{\Sigma})$ as the extended matrix 
obtained by inserting $k$ all-zero columns and $k$ all-zero rows after the 
first $n$ columns and rows of $\boldsymbol{\Sigma}$, respectively.  

\begin{theorem} \label{theorem2}
Let $\boldsymbol{Q}_{2L_{\mathrm{t}}}=\boldsymbol{U}_{L_{\mathrm{t}},1/2}
\boldsymbol{\Sigma}_{2L_{\mathrm{t}}}\boldsymbol{U}_{L_{\mathrm{t}},1/2}^{\mathrm{H}}$ and  
\begin{equation} \label{Q}
\boldsymbol{Q}_{M}=\boldsymbol{U}_{L_{\mathrm{t}},\Delta_{\mathrm{t}}}
\boldsymbol{E}_{L_{\mathrm{t}}+1,M-(2L_{\mathrm{t}}+1)}(\boldsymbol{\Sigma}_{2L_{\mathrm{t}}+1})
\boldsymbol{U}_{L_{\mathrm{t}},\Delta_{\mathrm{t}}}^{\mathrm{H}}. 
\end{equation}
Assume that 
\begin{equation} 
\mathrm{Tr}\left\{
 \left(
  \boldsymbol{E}_{L_{\mathrm{t}},1}(\boldsymbol{\Sigma}_{2L_{\mathrm{t}}})
  - \boldsymbol{\Sigma}_{2L_{\mathrm{t}}+1}
 \right)^{2}
\right\}\to0 
\end{equation}
holds as $L_{\mathrm{t}}\to\infty$. 
Under Assumptions~\ref{assumption1} and \ref{assumption2},  
for fixed SNR $\tilde{\gamma}>0$ and load $\alpha>0$ 
\begin{equation} \label{capacity_difference}
\frac{
 |C_{\mathrm{opt}}(\boldsymbol{Q}_{2L_{\mathrm{t}}};\tilde{\gamma},
  \boldsymbol{G}_{1/2,1/2})
 - C_{\mathrm{opt}}(\boldsymbol{Q}_{M};\tilde{\gamma}, 
 \tilde{\boldsymbol{G}}_{\Delta_{\mathrm{t}}, \Delta_{\mathrm{r}}})|
}{2\min\{L_{\mathrm{t}},L_{\mathrm{r}}\}}\to0 
\end{equation}
holds uniformly for all antenna separations 
$\Delta_{\mathrm{t}}, \Delta_{\mathrm{r}}\leq1/2$, 
covariance matrices $\boldsymbol{\Sigma}_{2L_{\mathrm{t}}}
\in\mathfrak{S}_{2L_{\mathrm{t}}}$, 
$\boldsymbol{\Sigma}_{2L_{\mathrm{t}}+1}\in\mathfrak{S}_{2L_{\mathrm{t}}+1}$, 
and all channel instances $\mathcal{C}\in\mathfrak{C}$ 
in the large-system limit. 
\end{theorem}
\begin{IEEEproof}
See a long version of this paper~\cite{Takeuchi16}. 
\end{IEEEproof}

Combining Theorems~\ref{theorem1} and \ref{theorem2} implies that 
the normalized capacity for the critically spaced case coincides with  
that for the densely spaced case in the large-system limit. Thus, 
there are no points in considering the densely spaced case, as long as 
Gaussian signaling is used. To the best of author's knowledge, the two 
theorems are the first results for finite SNRs, while the optimality of 
the critically spaced case was proved in the high SNR regime~\cite{Poon06}. 

\subsection{Non-Gaussian Signaling}
For a square root $\boldsymbol{Q}_{M}^{1/2}$ of the input covariance matrix 
$\boldsymbol{Q}_{M}$, we next investigate the precoded QPSK scheme 
$\boldsymbol{x}=\boldsymbol{Q}_{M}^{1/2}\boldsymbol{b}$, in which 
$\boldsymbol{b}=(b_{0},\ldots,b_{M-1})^{\mathrm{T}}$ has independent QPSK 
elements with unit power. The achievable rate 
$C(\boldsymbol{Q}_{M};\tilde{\gamma}, 
\boldsymbol{G}_{\Delta_{\mathrm{t}},\Delta_{\mathrm{r}}})$ of the precoded QPSK 
scheme is defined as 
\begin{equation} \label{QPSK_capacity}
C(\boldsymbol{Q}_{M};\tilde{\gamma}, 
\boldsymbol{G}_{\Delta_{\mathrm{t}},\Delta_{\mathrm{r}}}) 
= I(\boldsymbol{b};\boldsymbol{y}), 
\end{equation}
where the effective MIMO channel is given by 
\begin{equation} \label{effective_channel}
\boldsymbol{y} 
= \sqrt{\tilde{\gamma}}\boldsymbol{A}\boldsymbol{b} + \boldsymbol{w}, 
\end{equation}
with 
\begin{equation} \label{effective_channel_matrix}
\boldsymbol{A} 
= \boldsymbol{U}_{L_{\mathrm{r}},\Delta_{\mathrm{r}}}
\boldsymbol{G}_{\Delta_{\mathrm{t}},\Delta_{\mathrm{r}}}
\boldsymbol{U}_{L_{\mathrm{t}},\Delta_{\mathrm{t}}}^{\mathrm{H}}
\boldsymbol{Q}_{M}^{1/2}. 
\end{equation}

\begin{theorem} \label{theorem3}
Consider the input covariance matrix $\boldsymbol{Q}_{M}$ given by (\ref{Q}).  
Under Assumptions~\ref{assumption1} and \ref{assumption2}, 
for fixed SNR $\tilde{\gamma}>0$ and load $\alpha>0$ 
\begin{equation} \label{QPSK_capacity_dif}
\frac{|C_{\mathrm{opt}}(\boldsymbol{Q}_{M};
\tilde{\gamma},\boldsymbol{G}_{\Delta_{\mathrm{t}},\Delta_{\mathrm{r}}})  
-C(\boldsymbol{Q}_{M};\tilde{\gamma},
\boldsymbol{G}_{\Delta_{\mathrm{t}},\Delta_{\mathrm{r}}})|}
{2\mathrm{min}\{L_{\mathrm{t}},L_{\mathrm{r}}\}}\to0
\end{equation}
holds uniformly for all receive antenna separations 
$\Delta_{\mathrm{r}}\in(0,1/2]$, covariance matrices 
$\boldsymbol{\Sigma}_{2L_{\mathrm{t}}+1}\in\mathfrak{S}_{2L_{\mathrm{t}}+1}$, 
and all channel instances $\mathcal{C}\in\mathfrak{C}$ 
in the dense limit $\Delta_{\mathrm{t}}\to0$ after taking the large-system 
limit. 
\end{theorem}
\begin{IEEEproof}
See Section~\ref{sec4}. 
\end{IEEEproof}

Theorem~\ref{theorem3} implies that the precoded QPSK scheme can achieve 
the normalized capacity of the massive MIMO channel as the transmit antenna 
separation $\Delta_{\mathrm{t}}$ tends to zero. The result is analogous to 
the optimality of QPSK FTN signaling over the AWGN channel as the symbol 
period tends to zero~\cite{Yoo10}.  

\section{Proof of Theorem~\ref{theorem3}} \label{sec4}
\subsection{Outline} 
The proof strategy in \cite{Yoo10} is re-organized to prove 
Theorem~\ref{theorem3}. The proof consists of three steps. 
In a first step, a lower bound on the achievable 
rate~(\ref{QPSK_capacity}) is derived on the basis of the linear 
minimum mean-square error (LMMSE) receiver with successive interference 
cancellation (SIC). 

In a second step, we prove that the 
signal-to-interference-plus-noise ratio (SINR) in each stage of SIC tends 
to zero in the dense limit after taking the large-system limit. Then,  
the interference-plus-noise is replaced with a circularly symmetric 
complex Gaussian (CSCG) random variable 
by using the fact that, when QPSK is used, the worst-case additive noise 
in the low SINR regime is Gaussian~\cite{Chan71}. 

In the last step, we utilize the second-order optimality~\cite{Verdu02} of 
QPSK for the AWGN channel in the low SINR regime to replace the data symbols 
by optimal Gaussian data symbols. Theorem~\ref{theorem3} follows 
from the optimality of the LMMSE-SIC for Gaussian signaling.  

\subsection{LMMSE-SIC}
Using the chain rule for the mutual information~(\ref{QPSK_capacity}) yields  
\begin{equation} \label{QPSK_capacity1}
I(\boldsymbol{b};\boldsymbol{y}) 
= \sum_{m=0}^{M-1}I(b_{m};\boldsymbol{y}|b_{0},\ldots,b_{m-1}). 
\end{equation}
We consider the LMMSE estimator $\hat{b}_{m}$ of $b_{m}$ based on the known 
information $\boldsymbol{y}$ and $\{b_{m'}| m'\in[0:m)\}$ to obtain 
the lower bound 
\begin{equation} \label{LMMSE_Bound}
I(b_{m};\boldsymbol{y}|b_{0},\ldots,b_{m-1}) 
\geq I(b_{m};\hat{b}_{m}|b_{0},\ldots,b_{m-1}). 
\end{equation}
In (\ref{LMMSE_Bound}), the LMMSE estimator 
$\hat{b}_{m}$~\cite[Eq.~(8.66)]{Tse05} 
is given by 
\begin{equation} \label{LMMSE}
\hat{b}_{m} 
= \sqrt{\tilde{\gamma}}\boldsymbol{a}_{m}^{\mathrm{H}}\boldsymbol{\Xi}_{m}\left(
 \boldsymbol{y} - \sqrt{\tilde{\gamma}}\sum_{m'=0}^{m-1}\boldsymbol{a}_{m'}b_{m'} 
\right), 
\end{equation}
with 
\begin{equation} \label{MSE}
\boldsymbol{\Xi}_{m} 
= \left(
 \boldsymbol{I} + \tilde{\gamma}\sum_{m'=m+1}^{M-1}\boldsymbol{a}_{m'}
 \boldsymbol{a}_{m'}^{\mathrm{H}}
\right)^{-1}, 
\end{equation}
where $\boldsymbol{a}_{m}\in\mathbb{C}^{N}$ denotes the $m$th 
column vector of the effective channel 
matrix~(\ref{effective_channel_matrix}). 

Let us derive an explicit expression of the lower 
bound~(\ref{LMMSE_Bound}). Substituting (\ref{effective_channel}) into the 
LMMSE estimator~(\ref{LMMSE}) yields 
\begin{equation}
\frac{1}{\sqrt{\rho_{m}}}\hat{b}_{m} = \sqrt{\rho_{m}}b_{m} + v_{m},
\label{LMMSE_output}
\end{equation}
with
\begin{equation} \label{SINR} 
\rho_{m} 
= \tilde{\gamma}
\boldsymbol{a}_{m}^{\mathrm{H}}\boldsymbol{\Xi}_{m}\boldsymbol{a}_{m}. 
\end{equation}
In (\ref{LMMSE_output}), the interference-plus-noise $v_{m}\in\mathbb{C}$ is 
given by 
\begin{equation}
\sqrt{\rho_{m}}v_{m} =
\tilde{\gamma}\sum_{m'=m+1}^{M-1}\boldsymbol{a}_{m}^{H}\boldsymbol{\Xi}_{m}
\boldsymbol{a}_{m'}b_{m'} 
+ \sqrt{\tilde{\gamma}}
\boldsymbol{a}_{m}^{\mathrm{H}}\boldsymbol{\Xi}_{m}\boldsymbol{w}. 
\end{equation}
Since $v_{m}$ has unit variance from (\ref{MSE}), 
$\rho_{m}$ is regarded as the SINR for the LMMSE estimator $\hat{b}_{m}$. 
From (\ref{LMMSE_output}), we obtain 
\begin{equation} \label{LMMSE_Bound2}
I(b_{m};\hat{b}_{m}|b_{0},\ldots,b_{m-1}) = I(b_{m};\sqrt{\rho_{m}}b_{m}+v_{m}). 
\end{equation} 

\subsection{Worse-Cast Additive Noise}
\begin{lemma} \label{lemma1} 
Consider the input covariance matrix $\boldsymbol{Q}_{M}$ given by (\ref{Q}), 
and fix SNR $\tilde{\gamma}>0$ and load $\alpha>0$. 
Under Assumptions~\ref{assumption1} and \ref{assumption2}, 
there exists some constant $A_{\alpha}>0$ such that 
the multiuser efficiency $\rho_{m}/(\Delta_{\mathrm{t}}\tilde{\gamma})$ 
normalized by $\Delta_{\mathrm{t}}$ is bounded from above by $A_{\alpha}$ 
for all $\Delta_{\mathrm{t}}, \Delta_{\mathrm{r}}\in(0,1/2]$, 
covariance matrices $\boldsymbol{\Sigma}_{2L_{\mathrm{t}}+1}\in
\mathfrak{S}_{2L_{\mathrm{t}}+1}$, 
channel instances $\mathcal{C}\in\mathfrak{C}$, and all $m\in[0: M)$ 
in the large-system limit. 
\end{lemma}
\begin{IEEEproof}
Since the maximum eigenvalue of (\ref{MSE}) is bounded from above by $1$,  
we have an upper bound on the SINR (\ref{SINR}), 
\begin{equation}
\frac{\rho_{m}}{\tilde{\gamma}} 
= \frac{\boldsymbol{a}_{m}^{\mathrm{H}}\boldsymbol{\Xi}_{m}
\boldsymbol{a}_{m}}{\|\boldsymbol{a}_{m}\|^{2}}
\|\boldsymbol{a}_{m}\|^{2}
< \left\|
 \boldsymbol{G}_{\Delta_{\mathrm{t}},\Delta_{\mathrm{r}}}
 \boldsymbol{U}_{L_{\mathrm{t}},\Delta_{\mathrm{t}}}^{\mathrm{H}}
 \boldsymbol{Q}_{M}^{1/2}\boldsymbol{e}_{M,m}
\right\|^{2}, 
\label{SINR_bound}
\end{equation}
where $\boldsymbol{e}_{M,m}$ is the $m$th column of $\boldsymbol{I}_{M}$. 
Repeating the same argument yields 
\begin{equation}
\frac{\rho_{m}}{\tilde{\gamma}} 
< \sigma_{\mathrm{max}}^{2}
\left\|
 (2\mathrm{min}\{L_{\mathrm{t}}, L_{\mathrm{r}}\})^{1/2}
 \boldsymbol{Q}_{M}^{1/2}\boldsymbol{e}_{M,m}
\right\|^{2}, 
\end{equation}
where the maximum singular value $\sigma_{\mathrm{max}}>0$ of the 
channel matrix $\mathrm{min}\{2L_{\mathrm{t}}, 2L_{\mathrm{r}}\}^{-1/2}\boldsymbol{G}_{\Delta_{\mathrm{t}},\Delta_{\mathrm{r}}}$ 
is uniformly bounded from Assumption~\ref{assumption1}.  

Let $\tilde{\boldsymbol{U}}_{L_{\mathrm{t}},\Delta_{\mathrm{t}}}$ denote the 
$M\times(2L_{\mathrm{t}}+1)$ matrix obtained by eliminating 
the columns $m\notin\mathcal{M}$ from 
$\boldsymbol{U}_{L_{\mathrm{t}},\Delta_{\mathrm{t}}}$. 
From the definition~(\ref{Q}) of $\boldsymbol{Q}_{M}$, we have 
\begin{equation}
\frac{\rho_{m}}{\sigma_{\mathrm{max}}^{2}\tilde{\gamma}} 
< \mathrm{min}\{1, \alpha^{-1}\}\lambda_{\mathrm{max}}
\|\tilde{\boldsymbol{U}}_{L_{\mathrm{t}},\Delta_{\mathrm{t}}}^{\mathrm{H}}
\boldsymbol{e}_{M,m}\|^{2}, 
\end{equation}
where the maximum eigenvalue 
$\lambda_{\mathrm{max}}>0$ of 
$2L_{\mathrm{t}}\boldsymbol{\Sigma}_{2L_{\mathrm{t}}+1}$ 
is uniformly bounded from Assumption~\ref{assumption2}. 
Since the $m$th column 
of $\boldsymbol{U}_{L_{\mathrm{t}},\Delta_{\mathrm{t}}}$ is equal to 
$\boldsymbol{s}_{L_{\mathrm{t}},\Delta_{\mathrm{t}}}(m/L_{\mathrm{t}})$ given by 
(\ref{signature}), we have 
\begin{IEEEeqnarray}{rl} 
\|\tilde{\boldsymbol{U}}_{L_{\mathrm{t}},\Delta_{\mathrm{t}}}^{\mathrm{H}}
\boldsymbol{e}_{M,m}\|^{2}  
=& \frac{\Delta_{\mathrm{t}}}{L_{\mathrm{t}}}\sum_{m'\in\mathcal{M}}
|e^{2\pi jm\Delta_{\mathrm{t}}m'/L_{\mathrm{t}}}|^{2} 
\nonumber \\ 
=& \left(
 2 + L_{\mathrm{t}}^{-1}
\right)\Delta_{\mathrm{t}}.  
\end{IEEEeqnarray}
Thus, Lemma~\ref{lemma1} holds. 
\end{IEEEproof}

A further lower bound on (\ref{LMMSE_Bound2}) is derived by using the fact 
that Gaussian noise is the worst-case noise for the additive noise 
channel with QPSK in the low SINR regime~\cite{Chan71}\footnote{
Although the real additive noise channel with binary phase-shift keying was 
considered in \cite{Chan71}, a generalization to the QPSK case is 
straightforward.}. 
\begin{equation} \label{worst_case}
I(b_{m};\sqrt{\rho_{m}}b_{m}+v_{m}) 
\geq I(b_{m};\sqrt{\rho_{m}}b_{m}+v_{m}^{\mathrm{G}}), 
\end{equation}
with $v_{m}^{\mathrm{G}}\sim\mathcal{CN}(0,1)$, uniformly for all 
$\Delta_{\mathrm{r}}\in(0,1/2]$, covariance matrices 
$\boldsymbol{\Sigma}_{2L_{\mathrm{t}}+1}\in\mathfrak{S}_{2L_{\mathrm{t}}+1}$, and all 
channel instances $\mathcal{C}\in\mathfrak{C}$ in the dense limit 
$\Delta_{\mathrm{t}}\to0$ after taking the large-system limit.  
Applying (\ref{LMMSE_Bound}), (\ref{LMMSE_Bound2}), and (\ref{worst_case}) to 
(\ref{QPSK_capacity1}) yields 
\begin{equation} \label{QPSK_capacity2}
\frac{1}{M}I(\boldsymbol{b};\boldsymbol{y}) 
\geq \frac{1}{M}\sum_{m=0}^{M-1}I(b_{m};\sqrt{\rho_{m}}b_{m} + v_{m}^{\mathrm{G}})
\end{equation} 
in the dense limit after taking the large-system limit. 

\subsection{Second-Order Optimality of QPSK} 
We use the second-order optimality of the QPSK symbol $b_{m}$ to evaluate the 
mutual information $I(b_{m};\sqrt{\rho_{m}}b_{m}+v_{m}^{\mathrm{G}})$. 
Let $b_{m}^{\mathrm{G}}\sim\mathcal{CN}(0,1)$ denote a CSCG data symbol 
with unit power. Since QPSK is second-order optimal for the AWGN 
channel~\cite{Verdu02}, we have 
\begin{equation} \label{SO_optimality}
\frac{
 |I(b_{m};\sqrt{\rho_{m}}b_{m}+v_{m}^{\mathrm{G}}) 
 - I(b_{m}^{\mathrm{G}};\sqrt{\rho_{m}}b_{m}^{\mathrm{G}}+v_{m}^{\mathrm{G}})|
}{\rho_{m}^{2}} \to 0
\end{equation} 
uniformly for all $\Delta_{\mathrm{r}}\in(0,1/2]$, 
covariance matrices 
$\boldsymbol{\Sigma}_{2L_{\mathrm{t}}+1}\in\mathfrak{S}_{2L_{\mathrm{t}}+1}$, 
and all channel instances $\mathcal{C}\in\mathfrak{C}$ in the dense limit 
$\Delta_{\mathrm{t}}\to0$ after taking the large-system limit. 

Applying (\ref{SO_optimality}) to (\ref{QPSK_capacity2}), 
from Lemma~\ref{lemma1} we find 
\begin{equation}
\frac{1}{M}I(\boldsymbol{b};\boldsymbol{y}) 
\geq \frac{1}{M}C_{\mathrm{opt}}(\boldsymbol{Q}_{M};
\tilde{\gamma}, \boldsymbol{G}_{\Delta_{\mathrm{t}},\Delta_{\mathrm{r}}}) 
+ o(\Delta_{\mathrm{t}}^{2}),  
\label{QPSK_capacity3}
\end{equation}
in the dense limit $\Delta_{\mathrm{t}}\to0$ after taking the large-system 
limit. In the derivation of (\ref{QPSK_capacity3}), 
we have used the fact that the LMMSE-SIC for Gaussian signaling achieves 
the constrained capacity 
$C_{\mathrm{opt}}(\boldsymbol{Q}_{M};
\tilde{\gamma}, \boldsymbol{G}_{\Delta_{\mathrm{t}},\Delta_{\mathrm{r}}})$ 
given by (\ref{capacity})~\cite{Tse05}. 
Dividing both sides by $\Delta_{\mathrm{t}}$, we have 
\begin{equation} \label{QPSK_capacity4}
\frac{L_{\mathrm{t}}}{2\mathrm{min}\{L_{\mathrm{t}},L_{\mathrm{r}}\}}
\frac{I(\boldsymbol{b};\boldsymbol{y}) 
- C_{\mathrm{opt}}(\boldsymbol{Q}_{M};
\tilde{\gamma}, \boldsymbol{G}_{\Delta_{\mathrm{t}},\Delta_{\mathrm{r}}})}
{L_{\mathrm{t}}}\geq0
\end{equation}
in the dense limit $\Delta_{\mathrm{t}}\to0$ after taking the large-system 
limit. Since the upper bound 
$C_{\mathrm{opt}}(\boldsymbol{Q}_{M};
\tilde{\gamma}, \boldsymbol{G}_{\Delta_{\mathrm{t}},\Delta_{\mathrm{r}}})
-I(\boldsymbol{b};\boldsymbol{y})\geq0$ is trivial, we arrive at 
Theorem~\ref{theorem3}. 

\section{Numerical Simulation}
In numerical simulations, we use an independent $P$-path Rayleigh fading model 
to generate the channel matrix $\boldsymbol{G}_{\Delta_{\mathrm{t}},\Delta_{\mathrm{r}}}$ 
in the angular domain. In each discrete path, the attenuation is independently 
sampled from the CSCG distribution with variance $1/P$, while the departure 
and incident angles are independently sampled from the uniform distribution 
on $[0, 2\pi)$. The condition $P\geq2\mathrm{min}\{L_{\mathrm{t}}, L_{\mathrm{r}}\}$ 
is necessary for achieving the maximum spatial degrees of freedom. 

We focus on single-user MIMO downlink in which the transmitter has a larger 
antenna array than the receiver. For simplicity, the receiver is assumed to 
use the critical antenna separation $\Delta_{\mathrm{r}}=1/2$. 
Postulate that channel state information (CSI) is available at the receiver, 
while CSI is unknown to the transmitter. For the densely spaced case 
$\Delta_{\mathrm{t}}<1/2$, we consider the diagonal matrix 
$\boldsymbol{\Sigma}_{M}=
(2L_{\mathrm{t}}+1)^{-1}\mathrm{diag}\{\boldsymbol{1}_{L_{\mathrm{t}}+1},
\boldsymbol{0},\boldsymbol{1}_{L_{\mathrm{t}}}\}$ to generate the input covariance 
matrix $\boldsymbol{Q}_{M}$ given by (\ref{Q}), 
while $\boldsymbol{\Sigma}_{2L_{\mathrm{t}}}
=(2L_{\mathrm{t}})^{-1}\boldsymbol{I}_{2L_{\mathrm{t}}}$ is used for the critically 
spaced case $\Delta_{\mathrm{t}}=1/2$. 

\begin{figure}[t]
\begin{center}
\includegraphics[width=\hsize]{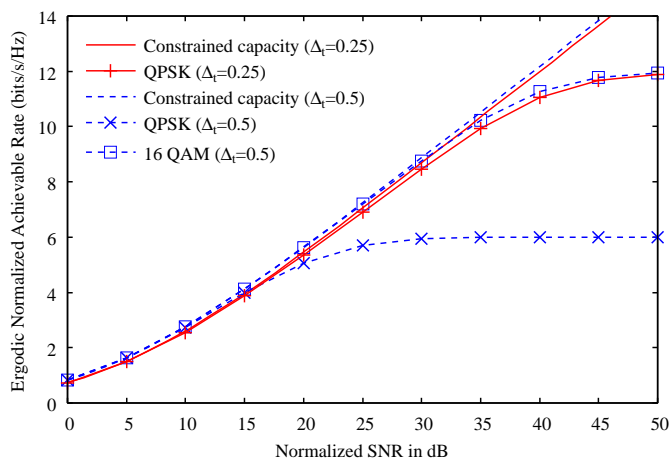}
\caption{
Ergodic achievable rate divided by 
$2\mathrm{min}\{L_{\mathrm{t}}, L_{\mathrm{r}}\}$ versus the normalized SNR 
$\tilde{\gamma}$ for $L_{\mathrm{t}}=3$, $L_{\mathrm{r}}=1$, 
$\Delta_{\mathrm{r}}=0.5$, and $P=4$. 
}
\label{fig2}
\end{center}
\end{figure}

Figure~\ref{fig2} shows the ergodic achievable rates versus the normalized 
SNR $\tilde{\gamma}=\gamma/(4\Delta_{\mathrm{t}}\Delta_{\mathrm{r}})$, in which 
$\gamma$ denotes the actual SNR. We find that the constrained capacity 
for the densely spaced case $\Delta_{\mathrm{t}}=1/4$ is slightly smaller than 
that for the critically spaced case $\Delta_{\mathrm{t}}=1/2$. 
$16$ quadrature amplitude modulation (QAM) and QPSK get closer toward the 
corresponding constrained capacity, respectively, as the normalized SNR 
decreases. The latter observation is consistent with the second-order 
optimality of the modulation schemes for the MIMO channel in the low SNR 
regime~\cite{Verdu02}. 

Unfortunately, QPSK for the densely spaced case is slightly inferior to 
$16$~QAM for the critically spaced case, while it is superior to QPSK 
for the critically spaced case in the high SNR regime. However, the gap 
between QPSK for the densely spaced case and $16$~QAM would shrink if 
larger antenna arrays were used, since Theorems~\ref{theorem1} and 
\ref{theorem2} imply that the gap in capacity tends to zero in the 
large-system limit. We conclude that using densely spaced transmit antennas 
is an alternative approach to increasing the achievable transmission rate, 
instead of using higher-order modulation.  

\balance

\section*{Acknowledgment}
The author was in part supported by the Grant-in-Aid for Young Scientists (A) 
(MEXT/JSPS KAKENHI Grant Number 26709029), Japan, and by the Grant-in-Aid 
for Exploratory Research (JSPS KAKENHI Grant Number 15K13987), Japan.




\bibliographystyle{IEEEtran}
\bibliography{IEEEabrv,kt-isit2016}
\end{document}